\documentclass[twocolumn,showpacs,preprintnumbers,amsmath,amssymb]{revtex4}
\input{epsf.tex}

\def\be{\begin{equation}}
\def\ee{\end{equation}}
\def\ba{\begin{array}}
\def\ea{\end{array}}
\def\bea{\begin{eqnarray}}
\def\eea{\end{eqnarray}}
\begin{document}
\draft
%%%%%%%%%%%%%%%%%%%%%%%%%%%%%%%%%%%%
\title {Disappearance of Transverse Flow in Central Collisions for
Heavier Nuclei}
%%%%%%%%%%%%%%%%%%%%%%%%%%%%%%%%
\author{Aman D. Sood,$^{1}$ Rajeev K. Puri,$^{1}$ and J\"org
Aichelin$^{2}$\\
\it $^1$Department of Physics, Panjab University, Chandigarh -160 014, India\\
\it $^2$SUBATECH, Ecole des Mines de Nantes, \it 4 rue Alfred Kastler, F-44070
Nantes Cedex, France}
%%%%%%%%%%%%%%%%%%%%%%%%
\begin{abstract}
Using the Quantum Molecular Dynamics model,
we analyze the disappearance of
flow in heavier colliding nuclei. A power law mass dependence
($\propto{\frac{1}{\sqrt{A}}}$) is obtained in
all cases. Our results are in excellent
agreement with experimental data which allows us to
predict the balance energy for
$^{238}$U + $^{238}$U collision around 37-39 MeV/nucleon.
\end{abstract}
\maketitle
Thirty years ago it was predicted by Scheid and Greiner \cite{Sch68} that
in heavy ion reactions the nuclei will be compressed and heated and that this
yields for non central reactions to in-plane flow $\langle p_x^{dir} \rangle$.
More than a
decade later, this conjecture was confirmed by the Plastic Ball group
\cite{pb}. In the following investigations it turned out that this in-plane flow
carries information on the nuclear equation of state \cite{ho}. If the nuclear
equation of state is stiffer, more compressional energy will be stored in
semi-central reactions and, when released, more in-plane flow will be given
to the nucleons.

The maximal density which is reached in a reaction depends on the beam
energy as well as on the system size. The lower the beam energy the less is the 
compression. At very low energies, the repulsive part of the nuclear equation
of state, which appears at densities above the normal nuclear matter
density, is not tested anymore and the nucleons feel only the attractive mean
field. A typical example is the deep inelastic reactions in which the two
nuclei rotate around a common center. This rotation creates in-plane flow as
well but in  opposite direction: Due to the common rotation the nucleons
stick together for a while and will be emitted into the direction opposite to 
the impact parameter whereas the in-plane flow which is caused by compression 
will be in the direction of the impact parameter.

There is a beam energy at which the in-plane flow disappears when changing from
the direction into that opposite to the impact parameter.
It has been shown in the simulation of heavy ion reactions that this beam
energy called balance energy, $E_{bal}$,
\cite{3a,1a,2a} depends on the nucleon-nucleon (nn) cross-section 
in the medium as well as on the potential \cite{1a,2a}. With the very recently 
measured $E_{bal}$ in $^{197}$Au + $^{197}$Au collisions \cite{4a} (earlier only estimated values
were available \cite{5a}),
a renewed interest has emerged in the field \cite{6a}. In addition to
the Au system, balance energies $E_{bal}$ of $^{12}$C + $^{12}$C \cite{7a},
$^{20}$Ne + $^{27}$Al \cite{7a}, $^{36}$Ar + $^{27}$Al \cite{8a},
$^{40}$Ar + $^{27}$Al \cite{9a}, $^{40}$Ar + $^{45}$Sc \cite{4a,7a,14a},
$^{40}$Ar + $^{51}$V \cite{2a,10a}, $^{64}$Zn + $^{27}$Al \cite{11a},
$^{40}$Ar + $^{58}$Ni \cite{6a}, $^{64}$Zn + $^{48}$Ti \cite{8a},
$^{58}$Ni + $^{58}$Ni \cite{4a,6a}, $^{64}$Zn + $^{58}$Ni \cite{8a},
$^{86}$Kr + $^{93}$Nb \cite{4a,7a},
$^{93}$Nb + $^{93}$Nb \cite{3a}, $^{129}$Xe + $^{118}$Sn \cite{6a}
and $^{139}$La + $^{139}$La \cite{3a} are also available.
It is worth mentioning that most of the above studies were for
the central collisions only. A few, however, also searched for the
impact parameter dependence of the balance energy \cite{4a,9a,14a,11a}.

Apart from the directed in-plane flow, differential as well as elliptic flow 
has also been predicted very recently \cite{12a}. 

The measurements of the balance energy over wide range of
system sizes provide an excellent opportunity to pin down the role
of  the mass dependence, where only preliminary studies \cite{4a,7a}
have been performed yet. These preliminary studies                                                
suggest a power law dependence $\propto{A^{\tau}}$ of the balance energy on
the mass number of the system. Interestingly, most of the theoretical studies 
are done within the Boltzmann-Uehling-Uhlenbeck (BUU) model
\cite{3a,1a,2a,4a,7a,9a,11a,12a,13a,15a,16a,17a,18a}. Some attempts, however,
also exist within the framework of Quantum
Molecular Dynamics (QMD) model \cite{14a,19a,soff95,21a}. 
Heavy systems are rather rarely analyzed in these approaches.

Our present aim is therefore to study the mass dependence of the balance 
energy in heavy colliding nuclei and to predict for the first time 
the disappearance of the collective in-plane flow in 
central $^{238}$U + $^{238}$U collision. We shall show that the mass
dependence of $E_{bal}$ for heavier nuclei scales approximately more as
$\frac{1}{\sqrt{A}}$ rather than as
$A^{-\frac{1}{3}}$ as has been suggested for light and medium colliding nuclei
\cite{7a}. The present study is made
within the framework of QMD model which is described in detail in refs.
\cite{19a,soff95,21a,22a,23a}.
%%%%%%%%%%%%%%%%%%%%%%%%%%%%%%%%%

%\section{The Model}
In the QMD model, each nucleon propagates under the
influence of mutual interactions. The propagation is governed by the
classical equations of motion:
%%%%%%%%%%%%%%%%%%%%%%%%%%%%%%%%%%
\begin{equation}
\dot{{\bf r}}_i~=~\frac{\partial H}{\partial{\bf p}_i};
~\dot{{\bf p}}_i~=~-\frac{\partial H}{\partial{\bf r}_i},
\end{equation}
%%%%%%%%%%%%%%%%%%%%%%%%%555
where H stands for the Hamiltonian which is given by:
%%%%%%%%%%%%%%%%%%%%%%%%%%%%%%%%%%%%%%5
\begin{equation}
H = \sum_i^{A} {\frac{{\bf p}_i^2}{2m_i}} + \sum_i^{A} ({V_i^{Skyrme} + V_i^{Yuk} +
V_i^{Coul}}).
\end{equation}
%%%%%%%%%%%%%%%%%%%%%%%%%%%%%%
Here $V_{i}^{Skyrme}$, $V_{i}^{Yuk}$ and $V_{i}^{Coul}$ are, respectively,
the Skyrme, Yukawa and Coulomb potentials. Momentum dependent
interactions are not important at these low beam energies as Zhou et al.
\cite{18a} have shown. It is worth mentioning that
both the soft and hard equations of state have been employed 
in the literature. Following \cite{3a,1a,9a,14a,11a,12a,17a,18a,19a,soff95,21a,24a},
we shall also use a hard equation of state.
For the nn cross-section  we chose an
isotropic and energy independent cross-section of 40 mb for the 
present analysis. This seems to us a reasonable choice in view of the
fact that cross-sections based on G-Matrix calculations differ widely and
because at this energy most of the collisions are 
Pauli blocked. Hence different cross-sections do not produce a large effect if
they are not too small.

Since we plan to study the heavier colliding nuclei, different nn
cross-sections should not have much effect \cite{4a}. 
Further,
it has been shown in ref. \cite{15a} that the nucleons in the present energy 
domain ($\leq$ 80 MeV/nucleon) collide with average $\sigma$=55 mb.
The isotropy of the cross-section also does not affect 
the reaction dynamics \cite{25a}. Similar assumptions were also made in
refs. \cite{1a,9a,11a,17a,21a,26a,27a}.
One should, however, keep in the mind that
different nn cross-sections may affect the dynamics in lighter systems.
%%%%%%%%%%%%%%%%%%%%%%%%%%%%%%%%%%%%%%%%

%\section{Results and Discussion}
Using the above description, we simulated the central
collisions with system mass A ($=A_{T}+A_{P}$; $A_{T}$ being the target
mass, and $A_{P}$ being the projectile mass)
$\geq175$. In particular, we simulated
$^{86}$Kr + $^{93}$Nb (b = 4.07 fm) \cite{7a}, $^{93}$Nb + $^{93}$Nb
(b = 3.104 fm) \cite{3a}, $^{129}$Xe + $^{118}$Sn (b = 0-3 fm) \cite{6a},
$^{139}$La + $^{139}$La (b = 3.549 fm) \cite{3a},
$^{197}$Au + $^{197}$Au (b = 2.5 fm) \cite{4a} and
$^{238}$U + $^{238}$U (b = 0-3 fm)
at incident energies between
30 MeV/nucleon and 80
MeV/nucleon at a step of 10 MeV. A straight line interpolation between steps
was used to calculate
the energy of vanishing flow $E_{bal}$. The reaction was followed till transverse flow
saturates which is close to 300 fm/c for heavier colliding nuclei whereas it
is $\leq200$ fm/c for lighter colliding nuclei. 
%It should, however, be kept
%in mind that impact parameter plays little role in heavy colliding nuclei
%\cite{4a}.
%%%%%%%%%%%%%%%%%%%%%%%%%%%%%%%%%%%%%%%??

In fig. 1, we display the average directed
transverse momentum $\langle p_{x}^{dir}\rangle$ defined as:
%%%%%%%%%%%%%%%%%%%%%%%%%%
\begin{equation}
\langle p_{x}^{dir}\rangle~=~\frac{1}{A} \sum_{i=1}^{A} sign\{Y(i)\}p_{x}(i),
\end{equation}
%%%%%%%%%%%%%%%%%%%%%%%%%%%%
where $Y(i)$ and $p_{x}(i)$ are, respectively, the rapidity  and the
transverse momentum of the $\it{ith}$ particle for the reactions  
$^{93}$Nb + $^{93}$Nb, $^{139}$La + $^{139}$La, $^{197}$Au + $^{197}$Au
and $^{238}$U + $^{238}$U. We see that the collective in-plane flow for
$^{93}$Nb + $^{93}$Nb changes sign between 
55-60 MeV/nucleon whereas it is already positive around 40 MeV/nucleon for
$^{238}$U + $^{238}$U system. Further, the saturation time 
increases with the size of the colliding nuclei.
%This indicates towards the low incident energies and ongoing interactions
%in heavier colliding nuclei. 
The early onset of the
flow in heavier colliding nuclei is due to the Coulomb forces that
are much stronger in heavier systems compared to lighter nuclei. In addition,
a large collision rate in heavier
systems also contributes towards the early onset of the flow.
%%%%%%%%%%%%%%%%%%%%%%%%%%%%%%%%%%%%%%%%%%%%%%%%%%%%%%%%%%%%

In fig. 2, we display
the energy of vanishing
flow ($E_{bal}$) as a function of the combined mass of the system.
As stated earlier, $E_{bal}$ is extracted using a straight line interpolation
between the calculated  values of the in-plane flow. Here
open squares represent our calculations whereas solid stars are the experimental
findings. The dotted and dash-double-dotted lines are, respectively, the power law
fits ($\propto{A^{\tau}}$) to the theoretical values including
$^{238}$U + $^{238}$U in one case and
excluding $^{238}$U + $^{238}$U in the other case. The fit to the experimental
points is represented by a solid line. All fits are obtained with
$\chi^{2}$ minimization. The values of
$E_{bal}$ (obtained with a stiff equation of state and 40 mb cross-section) are
very close to the experimentally measured $E_{bal}$.
The fit to the experimental data yields $\tau=-0.52451\pm0.06261$,
whereas that to 
theory yields $\tau=-0.53261\pm0.16373$. Once $^{238}$U + $^{238}$U is included,
the $\tau$ decreases to $-0.50872\pm0.10883$.
In other words, we observe a $\frac{1}{\sqrt{A}}$ dependence in the
$E_{bal}$. Similar dependence can also be obtained
with a least square fit.
Based on the above findings, we predict  $E_{bal}$ for the central
$^{238}$U + $^{238}$U reaction
around 37-39 MeV/nucleon (according to  power law fits, it is around
37-38 MeV/nucleon whereas QMD simulation predicts around 39 MeV/nucleon).
It is worth mentioning that most of the earlier  mass dependence
calculations \cite{4a,7a,18a} could not
reproduce the experimentally extracted slopes \cite{4a,7a}. Zhou
et al. \cite{18a} could reproduce the slope, however their analysis
was only done for lighter nuclei $\leq200$. Our calculations can reproduce the
experimentally extracted slope very closely, therefore, we can predict the
energy of vanishing flow in the $^{238}$U + $^{238}$U system. The large 
deviation of our $\tau$ value from the standard value ($\approx{-\frac{1}{3}}$) 
reflects the increasing importance of the Coulomb repulsion with the size
of the system, as noted in ref. \cite{4a}. There the
$\tau$ value is close to $-\frac{1}{3}$ for masses $\leq200$ \cite{7a,18a}, 
whereas
it increases to  $\approx{-0.45}$ when heavier systems like
$^{139}$La + $^{139}$La and $^{197}$Au + $^{197}$Au are
included. In the present analysis, we took only heavier nuclei ($A\geq175$).
Therefore the slope is steeper than the above cited values.
If one also takes the lighter nuclei into
consideration, our value also decreases to $-0.4$. In other words,
for lighter and medium
nuclei, the balance energy $E_{bal}$ emerges due to the interplay between
the mean field
and nucleon scattering. However, for heavier colliding nuclei, Coulomb
interaction is as well an important factor. It is of
interest to see the
contributions of the mean field (that includes the Coulomb interaction) and nn
collisions towards the transverse flow at the balance energy $E_{bal}$.
Following ref. \cite{19a},
we decomposed both these contributions in the simulations itself. At each time
step during the reaction, the momentum transferred due to the two- body
collisions and mutual mean field potential
is calculated separately using equation (3). 
The separation was done
at each simulated incident energy and a straight line interpolation was used.
The decomposition is plotted in fig. 3 as a function
of the total mass of the system. We find that the contribution of mean field
towards transverse momentum is negative whereas it turns repulsive for the
collision part. Both these contributions again obey a power law
behavior with $\tau=-0.70931\pm0.24234$.

%\section{Summary}
Summarizing, we present the
disappearance of flow in heavier colliding nuclei with a prediction of
balance energy for $^{238}$U + $^{238}$U around 37-39 MeV/nucleon. Our calculations
(with a stiff
equation of state and $\sigma=40$ mb) are in a very close agreement with
experimentally extracted values ($\tau_{th}=-0.53261\pm0.16373$;
$\tau_{expt}=-0.52451\pm0.06261$). 
Both these findings suggest a power law mass
dependence $\propto\frac{1}{\sqrt{A}}$. The contribution of the mean field
towards flow at $E_{bal}$ is negative whereas it is positive for the
collision part. Both contributions can be parameterized in terms of a
power law.\\
%%%%%%%%%%%%%%%%%%%%%%%%%%%%%%%%%%%%%%%%

{\it This work is supported by the grant (No. SP/S2/K-21/96)
from the Department of Science and Technology, Government of India.}
%%%%%%%%%%%%%%%%%%%%%%%%%%%%%%%%%%%%%%%%%%%%5

%%%%%%%%%%%%%%%%%%%%%%%%%%%%%%%%%%
\newpage
{\Large \bf Figure Captions}\\

{\bf FIG. 1.} The time evolution of $\langle p_{x}^{dir} \rangle$ for four
different reactions: (a) $^{93}$Nb + $^{93}$Nb, (b) $^{139}$La + $^{139}$La,
(c) $^{197}$Au + $^{197}$Au
and (d) $^{238}$U + $^{238}$U. Here a stiff equation of state along
with constant nn cross-section of 40 mb strength is used.\\
%%%%%%%%%%%%%%%%%%%%%%%%

{\bf FIG. 2.} The $E_{bal}$ as a function of the total mass of the system. Solid stars
are the experimental data whereas open squares are the present theoretical
results. The solid line is a $\chi^{2}$ minimization fit of power law
($\propto{A^{\tau}}$) for
experimental data whereas dash-double-dotted line is a fit for
the corresponding theoretical result. The theoretical fit
that includes $^{238}$U + $^{238}$U reaction is represented by dotted line.\\
%%%%%%%%%%%%%%%%%%%%%%%%%%%%%%%%%%%%

{\bf FIG. 3.} The decomposition of $\langle p_{x}^{dir} \rangle$ at $E_{bal}$
into mutual mean field part and collision part as a function of the system size.
The lines are the $\chi^{2}$ fit of power law $\propto{c.A^{\tau}}$.

\begin{thebibliography}{999}
{\small
\bibitem{Sch68} W. Scheid, R. Ligensa, and W. Greiner, Phys. Rev. Lett. {\bf 21,}
1479 (1968).
\bibitem{pb} H.A. Gustafsson et al., Phys. Rev. Lett. {\bf 52,} 1590 (1984).
\bibitem{ho} H. St\"ocker and W. Greiner, Phys. Rep. {\bf 137,} 277 (1986).

\bibitem{3a} D. Krofcheck {\it et al.}, Phys. Rev. C {\bf 46,} 1416 (1992).

\bibitem{1a} J.J. Molitoris and H. St\"ocker, Phys. Lett. B {\bf 162}, 47
(1985); G.F. Bertsch, W.G. Lynch, and M.B. Tsang, Phys. Lett. B {\bf 189,}
384 (1987).

\bibitem{2a} C.A. Ogilvie {\it et al.}, Phys. Rev. C {\bf 42,} R10 (1990).

\bibitem{4a} D.J. Magestro, W. Bauer, O. Bjarki, J.D. Crispin, M.L. Miller,
M.B. Tonjes, A.M. Vander Molen, G.D. Westfall, R. Pak, and E. Norbeck,
Phys. Rev. C {\bf 61,} 021602(R) (2000);
D.J. Magestro, W. Bauer, and G.D. Westfall, ibid.
{\bf 62,} 041603(R) (2000); G.D. Westfall, Nucl. Phys. {\bf A681,} 343c (2001).

\bibitem{5a} W.M. Zhang {\it et al.}, Phys. Rev. C {\bf 42,} R491 (1990);
M.D. Partlan {\it et al.}, Phys. Rev. Lett. {\bf 75,} 2100 (1995);
P. Crochet {\it et al.}, Nucl. Phys. {\bf A624,} 755 (1997).

\bibitem{6a} D. Cussol {\it et al.}, Phys. Rev. C {\bf 65,} 044604 (2002).

\bibitem{7a} G.D. Westfall {\it et al.}, Phys. Rev. Lett. {\bf 71,} 1986
(1993).

\bibitem{8a} A. Buta {\it et al.}, Nucl. Phys. {\bf A584,} 397 (1995).

\bibitem{9a} J.P. Sullivan {\it et al.}, Phys. Lett. B {\bf 249,} 8 (1990).

\bibitem{14a} R. Pak {\it et al.}, Phys. Rev. C {\bf 54,} 2457 (1996);
 R. Pak {\it et al.}, ibid. {\bf 53,}
R1469 (1996).

\bibitem{10a} D. Krofcheck {\it et al.}, Phys. Rev. C {\bf 43,} 350 (1991).

\bibitem{11a} Z.Y. He {\it et al.}, Nucl. Phys. {\bf A598,} 248 (1996).

\bibitem{12a} Y.M. Zheng, C.M. Ko, B.A. Li, and B. Zhang,
Phys. Rev. Lett. {\bf 83,} 2534 (1999);
B.A. Li and A.T. Sustich, ibid. {\bf 82,} 5004 (1999).

\bibitem{13a} R. Pak {\it et al.}, Phys. Rev. Lett. {\bf 78,} 1022 (1997).
         
\bibitem{15a} B.A. Li, Phys. Rev. C {\bf 48,} 2415 (1993).

\bibitem{16a} V. de la Mota, F. Sebille, M. Farine, B. Remaud, and P.
Schuck, Phys. Rev. C {\bf 46,} 677 (1992).

\bibitem{17a} H.M. Xu, Phys. Rev. Lett. {\bf 67,} 2769 (1991); H.M. Xu, Phys.
Rev. C {\bf 46,} R389 (1992).

\bibitem{18a} H. Zhou, Z. Li, and Y. Zhuo, Phys. Rev. C
{\bf 50,} R2664 (1994).

\bibitem{19a} E. Lehmann, A. Faessler, J. Zipprich, R.K. Puri, and
S.W. Huang, Z. Phys. A {\bf 355,} 55 (1996).

\bibitem{soff95} S. Soff, S.A. Bass, C. Hartnack, H. St\"ocker, and
W. Greiner, Phys. Rev. C {\bf 51,} 3320 (1995).

\bibitem{21a} S. Kumar, M.K. Sharma, R.K. Puri, K.P. Singh, and I.M. Govil,
Phys. Rev. C {\bf 58,} 3494 (1998).

\bibitem{22a} J. Aichelin, Phys. Rep. {\bf 202,} 233 (1991).

\bibitem{23a} C. Hartnack, R.K. Puri, J. Aichelin, J. Konopka, S.A. Bass,
H. St\"ocker, and W. Greiner, Eur. Phys. J. A {\bf 1,} 151 (1998).

\bibitem{24a} J.J. Molitoris and H. St\"ocker, Phys. Rev. C
{\bf 32,} 346 (1985).

\bibitem{25a} S. Kumar, R.K. Puri, and J. Aichelin, Phys. Rev. C {\bf 58,}
1618 (1998).

\bibitem{26a} H.W. Barz, J.P. Bondorf, D. Idier, and I.N. Mishustin, Phys.
Lett. B {\bf 382,} 343 (1996).

\bibitem{27a} C. Roy {\it et al.}, Z. Phys. A {\bf 358,} 73 (1997).}

\end{thebibliography}
\end{document}